\documentclass{JINST}
\usepackage{amsmath,amssymb,mathrsfs,graphicx, subfig}

\title{Investigation of gamma-ray sensitivity of neutron detectors based on thin converter films}

\author{A. Khaplanov$^{a,b}$\thanks{Corresponding author.}~,
F.~Piscitelli$^{b,c}$, J.-C.~Buffet$^b$, J.-F.~Clergeau$^b$, J.~Correa$^b$, P.~van Esch$^b$, M.~Ferraton$^b$, B.~Guerard$^b$, R.~Hall-Wilton$^a$\\
\llap{$^a$}European Spallation Source,\\
  P.O Box 176, SE-22100 Lund, Sweden\\
\llap{$^b$}Institute Laue Langevin,\\
  Rue Jules Horowitz, FR-38042 Grenoble, France\\
\llap{$^c$} University of Perugia, \\
Piazza Universit\`a 1, 06123 Perugia, Italy.\\
E-mail: \email{Anton.Khaplanov@esss.se}}

\abstract{Currently, many detector technologies for thermal neutron detection are in development in order to lower the demand for the rare $^3He$ gas. Gas detectors with solid thin film neutron converters readout by gas proportional counter method have been proposed as an appropriate choice for applications where large area coverage is necessary. In this paper, we investigate the probability for $\gamma$-rays to generate a false count in a neutron measurement. Simulated results are compared to measurement with a $^{10}B$ thin film prototype and a $^3He$ detector. It is demonstrated that equal $\gamma$-ray rejection to that of $^3He$ tubes is achieved with the new technology. The arguments and results presented here are also applicable to gas detectors with converters other than solid $^{10}B$ layers, such as $^6Li$ layers and $^{10}BF_3$ gas.}


\begin{document}

\section{Introduction}
\label{sec:intro}

In the field of neutron scattering science, detectors for thermalized neutrons are required for a wide variety of instruments. In many cases, large area coverage of tens of square meters is required~\cite{cite:in5, cite:cdr, cite:tdr}. Up until recently, $^3He$ has been available in sufficient quantities for such detectors. This is no longer the case~\cite{cite:he31, cite:he32, cite:he33}, and a number of research programs are now aiming to find technologies that would replace $^3He$~\cite{cite:karl, cite:icnd}, in particular in a large number of new instruments of the upcoming European Spallation Source (ESS). One promising technique uses thin films containing $^{10}B$ as the neutron converter, while retaining the gas proportional counter as the sensing medium. This approach has already been implemented in a number of ways: \cite{cite:khaplanov, cite:correa, cite:guerard, cite:lacy, cite:stefanescu, cite:klein, cite:henske}. Also $^6Li$ solid converter layer have been used~\cite{cite:nelson}. Other venues of research include scintillator and semiconductor detectors.

The sensitivity of a neutron detector to $\gamma$-rays is a very important characteristic, as it defines the best achievable signal-to-noise ratio. In a neutron scattering instrument, typically a large amount of thermal neutron collimation is used to spatially define the neutron beam. In many cases also a temporal definition is required and is achieved by placing choppers -- rotating absorbers with well-defined openings. Further absorbers are found in beam stops that terminate residual beam after interaction with a sample or beams that are unused. Neutron captures in the neutron guides also contribute to the $\gamma$-ray background. The majority of efficient neutron absorbers, in particular $^{10}B$, $^{113}Cd$, and $^{155, 157}Gd$, emit one or more $\gamma$-rays for each neutron captured. Considering that scattering cross sections of many samples tend to be relatively low, the neutron intensity that carries useful information can be many orders of magnitude lower than the flux of $\gamma$-rays. The requirement of a low gamma sensitivity is also crucial for neutron detectors used in the Homeland Security programs, since in this case a low neutron signal must be detectable in the presence of natural $\gamma$-ray background, and the system should not create alarms due to a change in $\gamma$ flux.

In this paper we investigate the mechanisms responsible for signals induced by $\gamma$-rays in a neutron detector. We do so based on our simulations and measurement with the Multi-Grid detector~\cite{cite:khaplanov, cite:correa, cite:guerard} currently developed by a collaboration between ILL and ESS. The detailed physical aspects have been studied using a GEANT4~\cite{cite:g4} simulation. It is demonstrated that the conclusions are valid generally for detectors where gas proportional cells are used for signal generation, regardless of the nature of the neutron converted, be it $^3He$ gas or films of solid of $^{10}B$ or $^6Li$.  We argue that with correctly chosen operation parameters, a thin-film detector has equally high level of $\gamma$-ray rejection as a conventional $^3He$ tube.

\section{Physical processes}

\subsection{Neutron Interactions}
\label{sec:neutrons}

Since thermal neutrons do not directly induce a charge signal when interacting with matter, a nuclear reaction is necessary to convert neutrons into charged particles. Particularly suitable reactions are those that result in heavy charged particles such as protons, alphas and other ions. Some of those with viable cross sections and products for detectors are listed below:

\begin{equation}\label{eq:reactions}
\begin{array}{ll}
n+{^{10}B} &\xrightarrow{94\%} {^7Li}(0.84 MeV) + {^4He}(1.47 MeV) + \gamma (0.48 MeV) \\
&\xrightarrow{6\%} {^7Li}(1.02 MeV) + {^4He}(1.78 MeV) \\
n+{^{6}Li} &\rightarrow{^3H}(2.74 MeV)+{^4He}(2.05 MeV) \\
n+{^{3}He} &\rightarrow{p}(0.57 MeV)+{^3H}(0.20 MeV) \\
\end{array}
\end{equation}

The resultant charged particles ionize the detection medium and give rise to a primary detector signal, that may further be amplified, as in the case of a proportional gas counter. The ranges of the particles above are on the order of a few $\mu m$ in solids and a few $mm$ in gasses. Further notable reactions are neutron captures in two isotopes of gadolinium, $^{155}Gd$ and $^{157}Gd$. These result in excited nuclei that emit internal conversion electrons and $\gamma$-rays with a range of energies. Here, the kinetic energy and range of the daughter nucleus can be neglected, while electron and $\gamma$-ray ranges are greater than those of the heavier charged particles of the same energy.

\subsection{Photon Interactions}
\label{sec:photons}

\begin{figure}[tbp] 
\centering
\includegraphics[width=1\textwidth]{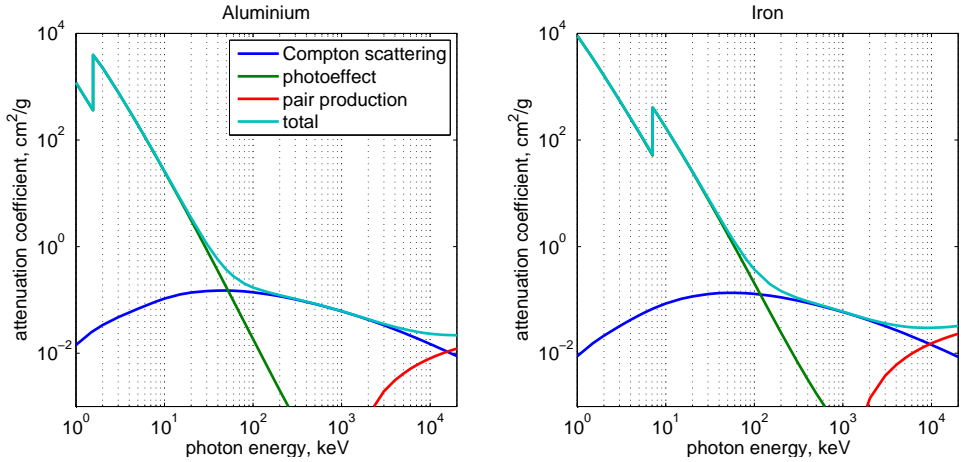}
\caption{Attenuation coefficients for $\gamma$-rays in aluminium and iron.}
\label{fig:photons}
\end{figure}

Similarly to the neutron, a photon can only be detected after converting into a charged particle. For the $\gamma$-ray range of energies that are likely to be found in a neutron scattering set-up, the relevant modes of interaction are photoelectric effect, Compton scattering and pair production. Figure~\ref{fig:photons} shows the contribution of each interaction mode as a function of $\gamma$-ray energy for aluminium and iron (common materials for mechanical detector parts). Low-energy photons typically will transfer all of their energy to an electron in a photoelectric interaction. Medium energies (corresponding to a majority of background $\gamma$-rays) Compton scatter, losing a part of their energy. Finally for the high-energy photons, pair production becomes dominant. 

Regardless of the interaction mode, an energetic electron (or positron) is created, that can generate a signal in the detector gas due to ionization of the medium. 

\subsection{Detector Response}
\label{sec:response}

While both neutron and $\gamma$ interactions can lead to ionization tracks in the detection medium, the energy deposition per unit length is very different. For a gas-filled detector, the range of protons and heavier particles is typically shorter than the dimensions of a gas cell, while the range of electrons is much larger for the same particle energies. Therefore the full energy of neutron conversion products that enter the gas can be measured in most events. On the other hand for electrons, the full energy is only measured for those electrons of the lowest energy, and only a limited fraction of the energy for more energetic electrons.

Considering the neutron signal, the major difference between the $^3He$ detector and thin film detectors is that in case of the gaseous neutron converter, the particles originate in the gas, and so have a high probability to deposit the full energy of at least one of the two particles in the gas. Therefore a neutron pulse height spectrum starts from the energy of the less-energetic conversion product, $^3H$ at $~\sim 200 keV$. On the other hand, photon interactions result in energies typically of a few tens on $keV$ and below. This leads to a very clear separation between energies deposited by neutrons and $\gamma$ and a simple lowest level discriminator threshold achieves an excellent rejection of gamma events. Note that this may no longer hold if charge loss effects, such as charge recombination, are present in the gas, or if the gas gain is high enough to saturate the signals.

For the case of neutron conversion in a film coating the inner walls of the detector, at most one particle can reach the gas after losing an arbitrary amount of energy while still in the solid. This means that a neutron event can have an arbitrarily low energy deposit in the gas, therefore mixing with the spectrum of $\gamma$ events. 

Photons may interact in both the solid walls as well as in the gas. Therefore the energy deposit can be arbitrarily low as well as limited in energy due to the finite gas cell size. It is important to understand that the change of the neutron converter material, from gas to solid, has minimal impact on the energy deposited by electrons produced by $\gamma$ interactions. Therefore an equally high level of gamma rejection can be achieved with any of these detectors. However, in the case of the solid converter, the neutron events that deposit an energy that may also be deposited by a $\gamma$-ray must be rejected.

%

\section{Multi-Grid Detector}
\label{sec:p11}

The Multi-Grid detector is one of the class of detectors for thermal neutrons where the conversion medium is a $^{10}B$-containing thin film. Several prototype detectors have been built by collaboration between the ESS and the ILL and tested at ILL. Thin films of enriched boron carbide, $^{10}B_4C$, have been developed and manufactured by Link{\"o}ping University~\cite{cite:hoglund}. The prototypes are described in detail elsewhere~\cite{cite:khaplanov, cite:correa, cite:guerard}; here a generic description is given common to all our prototypes. 

\begin{figure}[tbp] 
\centering
\includegraphics[width=.6\textwidth]{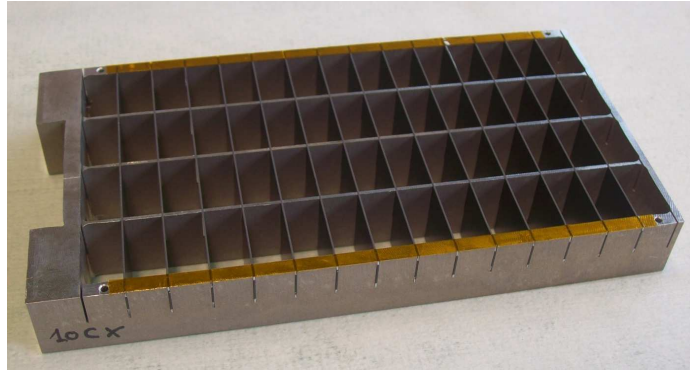}
\caption{A frame supporting $^{10}B_4C$ layers in the Multi-Grid detector. The dimensions of the unit cell are $10 \times 20 \times 20 mm$.}
\label{fig:frame}
\end{figure}

In order to achieve a neutron efficiency greater than 50\% at thermal neutron wavelength, 30 layers of of $^{10}B_4C$, each $1 \mu m$ thick, are placed orthogonally to the direction of the incoming neutrons. The layers are coated onto $0.5 mm$ thick $Al$ substrates which are then inserted into a $Al$ frame as shown in figure~\ref{fig:frame}. Several frames are then stacked so that the cells of the frames form long rectangular tubes, in which the surfaces traversed by neutrons are coated with the neutron converter, while those parallel to the neutron flux are usually not. Each such tube is equipped with an anode wire and acts as a proportional gas counter. A specific cell where an interaction occurred can be determined by recording coincidences between signals generated on wires and frames. 

\section{Simulation of Interactions}

A simulation of the Multi-Grid detector was made in GEANT4~\cite{cite:g4}. The simulation includes a variable number of cells corresponding to a single row of cells in a grid. A layer of neutron converter coats the cell faces orthogonal to the direction of incoming beam, the side walls are not coated. The cell is open on top and bottom. For the simulation presented here, $Al$ was used as the wall material and $^{10}B_4C$ the coating material. For the gas, $ArCO_2 (90/10)$ mixture and $CF_4$ were used. The energy of an event is defined as the sum off all interaction energies in the gas. A 2-$mm$ thick $Al$ entrance window is also included in the simulation in order to accurately predict the absolute detection efficiencies for photons (as well as for neutrons). 

\subsection{Treatment of Electromagnetic Interactions}

The functionality of GEANT4 classes describing the interactions of relevant to $\gamma$-rays -- photo effect, Compton scattering and electron energy loss -- has been extended in the recent years to be valid for low energies that are of interest in this work. The packages developed specifically for low energy electromagnetic interactions -- the Livermore Low-Energy model~\cite{cite:lowe} and the Penelope model \cite{cite:pene}, were however also tested. 

\begin{figure}[tbp] 
\centering
\includegraphics[width=1\textwidth]{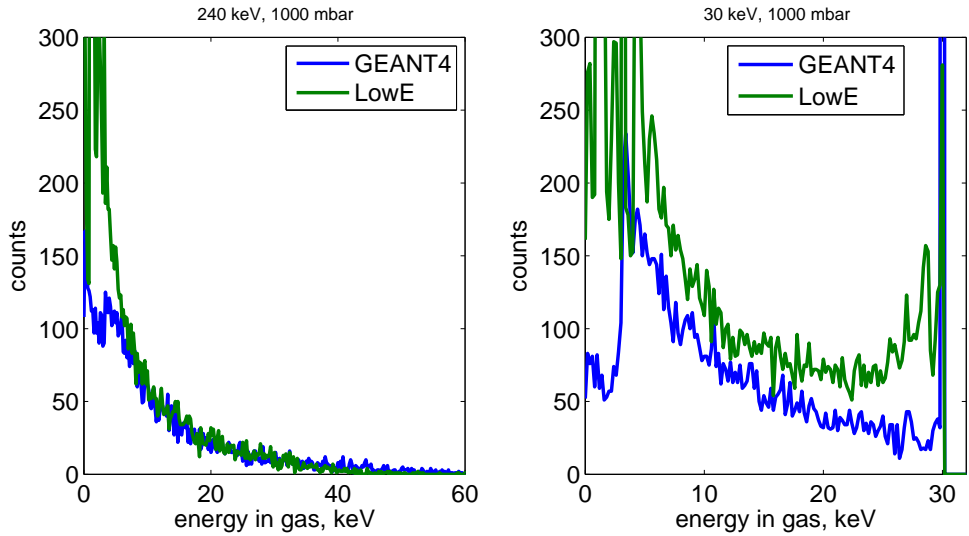}
\caption{Comparison of the treatment of low energy electromagnetic interactions by standard GEANT4 and Livermore \emph{LowE} classes for $\gamma$-ray energies of 240~keV (left) and 30~keV (right).}
\label{fig:lowe}
\end{figure}

While Low-Energy and Penelope models provided results indistinguishable from each other, the default GEANT4 model differed in that it did not generate secondary x-rays that follow an ejection of an electron through photo effect or Compton interaction. The inclusion of these x-rays will sometimes result in an event where the x-ray energy is absorbed in the gas even though no electron escapes the wall, or events where for an interaction in the gas, the x-ray escapes without being detected, thus reducing the measured energy. This is illustrated in figure~\ref{fig:lowe}. A full energy peak can only be found for very low energies, such as, $30~keV$. In the \emph{LowEnergy} treatment it becomes smeared towards lower energies due to x-ray escapes from the gas. The spectra also differ for energies of a few keV. The inclusion of these effects is clearly more correct, however, considering that the x-ray energies for the relevant materials are all below 3~keV -- far below any realistic discrimination threshold for neutrons -- the low energy models were generally not used (unless otherwise indicated) for the sake of faster processing times.

\subsection{Photon Interaction in Gas vs. Solid}

\begin{figure}[tbp] 
\centering
\includegraphics[width=1\textwidth]{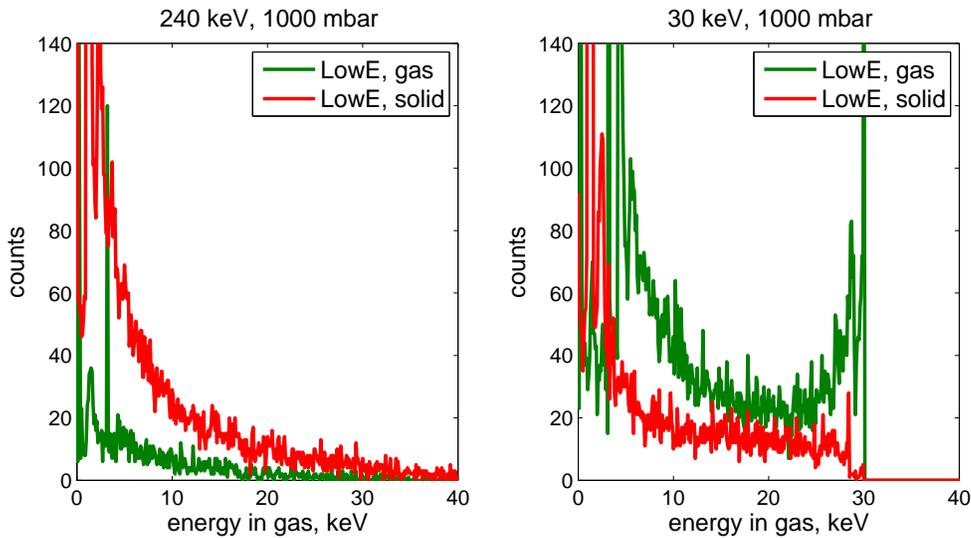}
\caption{Partial energy spectra due to interactions in gas and walls respectively for $\gamma$-ray energies of 240~keV (left) and 30~keV (right). \emph{LowEnergy} model was used for this simulation.}
\label{fig:gassolid}
\end{figure}

Interactions of photons with the gas are guaranteed to generate a signal. Those in the walls can only result in a signal if the electron reaches the gas, therefore only a few $\mu m$ on the surface of the wall will contribute. The relative effect of the two contributions can be seen in figure~\ref{fig:gassolid}. The two examples shown are for the cases where either component dominates -- gas interaction dominate for 30~keV and wall interaction dominate for 240~keV. The energy for which the two contributions are of equal magnitude is approximately 150~keV for $1 atm$ of $ArCO_2$ and $Al$ walls. 

\subsection{Effect of $\gamma$-ray Energy}

\begin{figure}[tbp] 
\centering
\includegraphics[width=.9\textwidth]{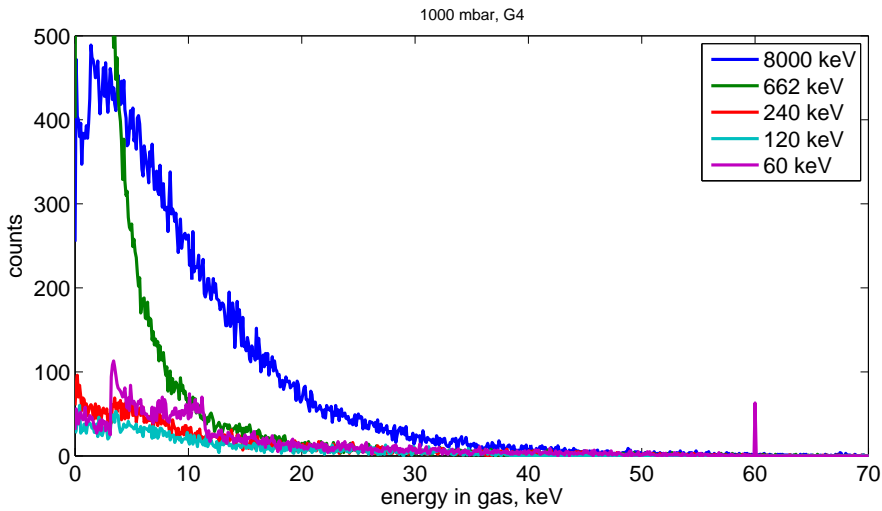}
\caption{Energy spectra for several $\gamma$-ray energies.}
\label{fig:energy}
\end{figure}

Interactions of $\gamma$-rays of several energies have been simulated and the energy deposited in gas determined. The resulting spectra are shown in figure~\ref{fig:energy}. The advantage of the gas detector becomes evident -- the deposited energies rarely exceed a few tens of keV regardless of the primary energy. In fact, the higher the $\gamma$-ray energy, the lower the average measured energy, because an electron of a higher energy has a lower energy loss per unit track length. Only for energies exceeding several MeV does the measured energy increase, however this effect is due to the electors that are energetic enough to traverse more than one cell. These cases can be rejected by the data acquisition system since signal multiplicity will be greater than one.

Integrating the number of events that exceed a threshold and normalizing to the total number of $\gamma$-rays simulated, yields an efficiency (or sensitivity) to gamma rays. This is tabulated in table~\ref{tab:efficiencies}. 

\begin{table}[tbp]
\caption{Percentage of $\gamma$-rays of a given energy that result in a detectable interaction for several thresholds in a simulation 1 million photons incident on a frame.}
\label{tab:efficiencies}
\smallskip
\centering
\begin{tabular}{lccccc}
\hline
Energy & \multicolumn{5}{c}{Threshold}\\
 \cline{2 - 6}
 & 1~keV & 10~keV & 30~keV & 60~keV & 100~keV \\ 
\hline
60~keV 		& 0.9815 \% 		& 0.4369 \%  		& 0.1000 \%  		& 0 \% 		& 0 \%\\
120~keV 		& 0.5297 \% 		& 0.2184 \% 		& 0.0351 \% 		& 0.0002 \% 	& 0 \% \\
240~keV 		& 0.8040 \% 		& 0.3143 \% 		& 0.0675 \% 		& 0.0015 \% 	& 0 \%\\
662~keV 		& 4.0161 \% 		& 0.4711 \% 		& 0.0711 \% 		& 0.0039 \%	 & 0 \%\\
8000~keV 	& 5.8324 \% 		& 2.1162 \% 		& 0.2199 \% 		& 0.0138 \% 	& $2 \times 10^{-4}$ \%\\
\hline
\end{tabular}
\end{table}

\subsection{Effect of Gas Pressure}

\begin{figure}[tbp] 
\centering
\includegraphics[width=1\textwidth]{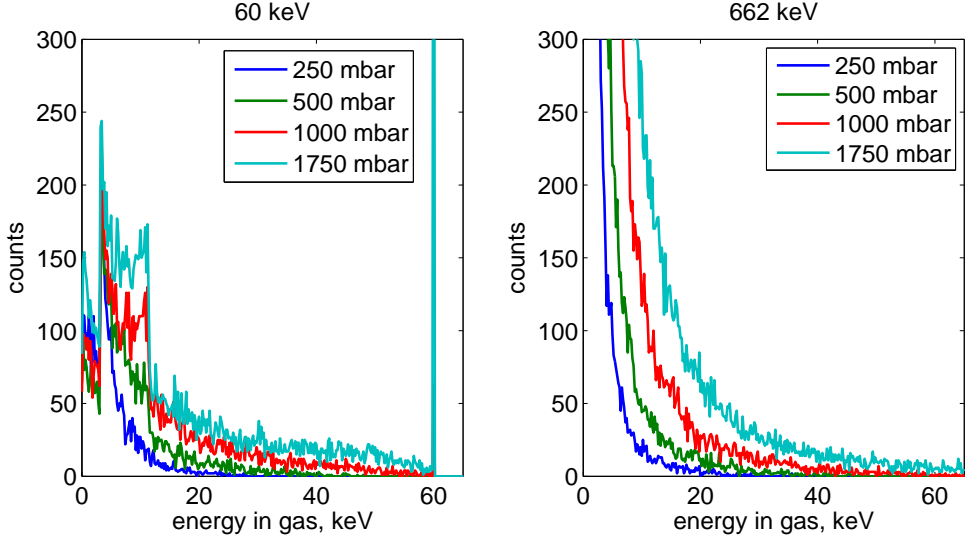}
\caption{Simulated energy spectra for 60 and 662~keV $\gamma$-ray sources for the $^{10}B$ detector filled with $ArCO_2 (90/10)$ gas mixture at different pressures, for $\gamma$-ray energies of 60~keV (left) and 662~keV (right).}
\label{fig:sim_pres}
\end{figure}

Pressure of the detecting gas has two effects on $\gamma$ detection. For higher energies, such as $662~keV$, the majority of detectable primary interactions occur in the walls of the gas cell. Photo or Compton electrons that enter the gas will lose more energy when traversing a gas of higher pressure. For low $\gamma$-ray energies, such as $60~keV$, the total cross section of the gas is more significant than the contribution of the solid layer from which electrons can reach the gas, therefore increasing the pressure will increase the total number of $\gamma$-rays converted. Furthermore, the specific energy loss in gas is also increased with pressure, resulting in a larger total energy deposit for a gas cell of a fixed size. Spectra for both energies at several gas pressures are shown in figure~\ref{fig:sim_pres}.

We see that for gas detectors with solid converters, it is advantageous to work at a low pressure from the point of view of gamma insensitivity. A low pressure is also preferable for detectors operated in vacuum vessels, as is the case in many neutron scattering instruments. Finally, a lower bias voltage can be used to achieve sufficient gas gain, reducing demands on electrical components. By contrast, a $^3He$ detector is typically operated with up to 10~bar of $^3He$ and 1-2~bar of a quenching gas such as $CF_4$. 

\section{Measurements}

Only measurements done for the purpose of characterizing the $\gamma$-ray sensitivity, are presented here. For a full characterization of our $^{10}B$ prototypes as well as the $^3He$ detector used throughout this work, see~\cite{cite:khaplanov, cite:correa}. 

\subsection{Measurements with $\gamma$ Sources}

\begin{figure}[tbp] 
\centering
\includegraphics[width=0.49\textwidth]{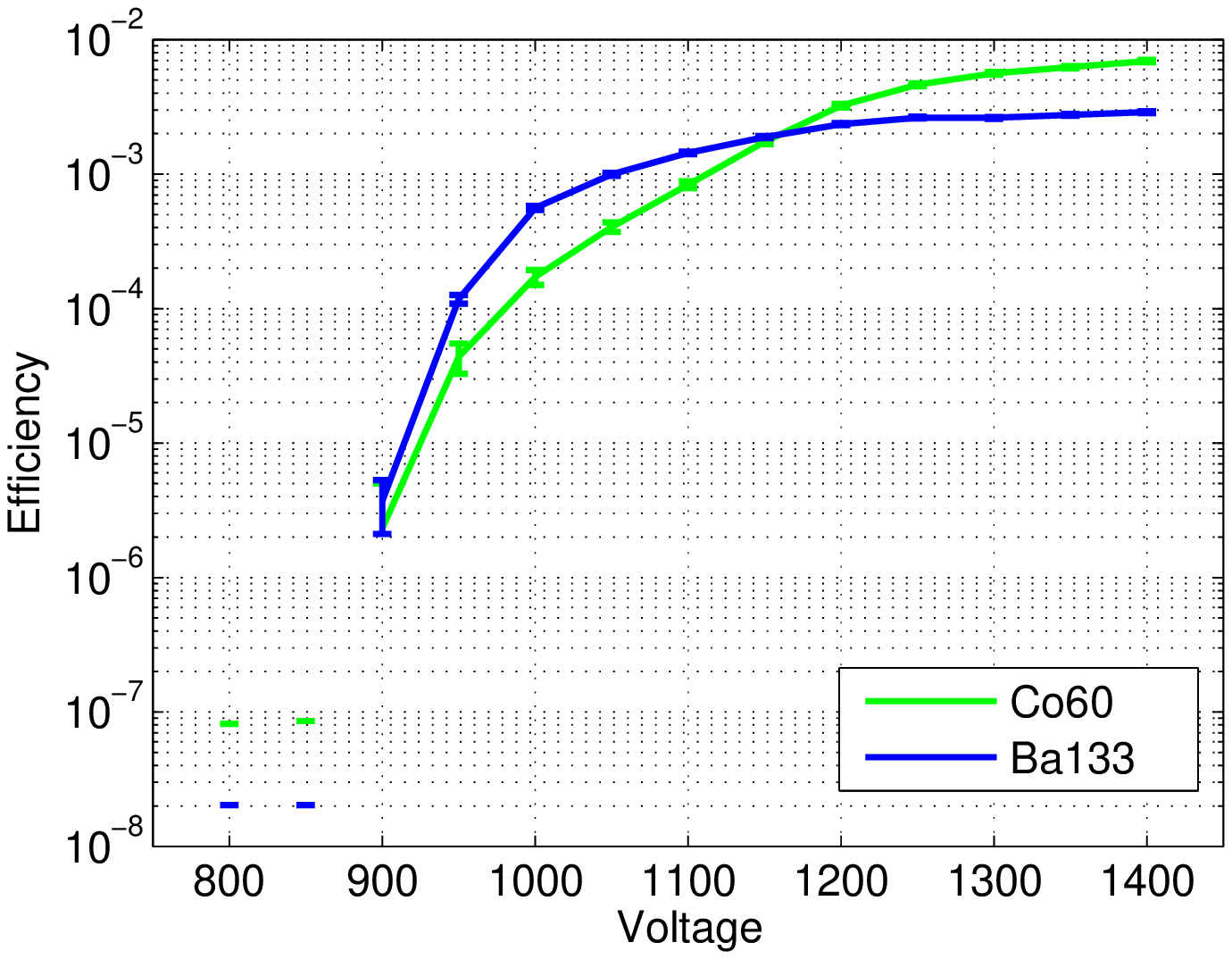}
\includegraphics[width=0.49\textwidth]{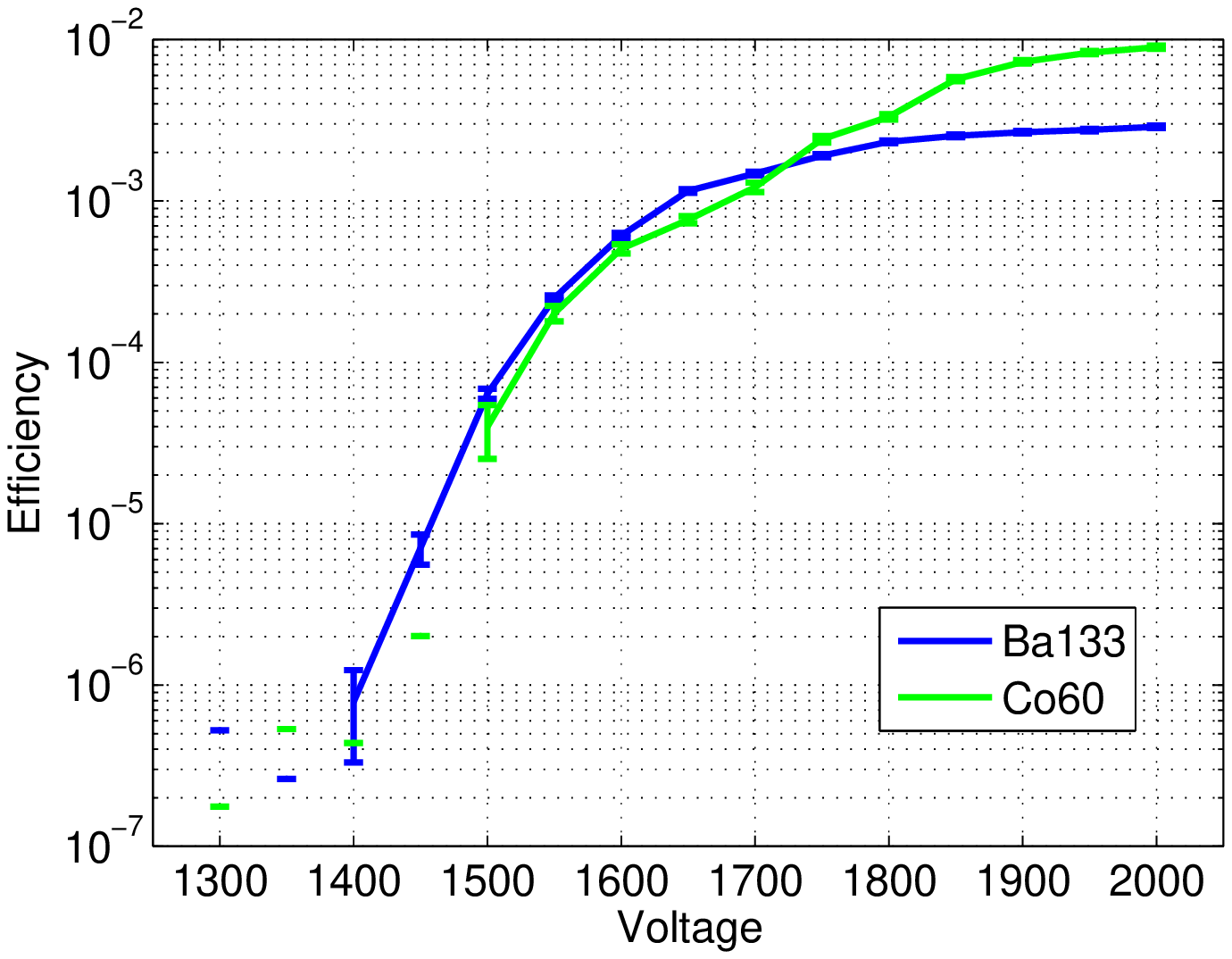}
\caption{Plateau measurements with the Multi-Grid $^{10}B$ detector (left) and a Multi-Tube $^3He$ detector (right) with $^{133}Ba$ (200~kBq) and $^{60}Co$ (25~kBq) $\gamma$-ray sources. Detection efficiency per tube is shown. The nominal operating voltages are 850V and 1350V respectively. The disconnected points at the left ends of the curves are upper limits, where no statistically-significant counts could be detected over background.}
\label{fig:sources}
\end{figure}

The response of the $^{10}B$ prototype to calibrated $\gamma$-ray sources has been measured. Counts and deposited energies were recorded for all cells of the detector. A similar set of measurements was performed with a Multi-Tube detector filled with 3 bars of $^3He$ and 1.5 bar of $CF_4$. This detector is composed of 37 hexagonal tubes with a $7 mm$ diameter arranged in a hexagonal formation. The results are presented in figures~\ref{fig:sources} and \ref{fig:sources2}, the $\gamma$-ray data for the sources used is shown in table~\ref{tab:sources}. Efficiencies or sensitivities to $\gamma$-rays are defined as the probability for a photon incident on a detector element (such as a tube) to result in an event. 

\begin{table}[tbp]
\caption{List of $\gamma$-rays emitted by the sources used in this work.}
\label{tab:sources}
\smallskip
\centering
\begin{tabular}{lcc}
\hline
  & Efficiency for threshold of\\
Source  & $\gamma$-ray, keV & intensity, \%\\ 
\hline
$^{133}Ba$ 	&	31		&	96.1\\
 			&	35	  	& 	17.3\\
			&	81		& 	32.9 \\
			& 	276		& 	7.2\\
			& 	303		& 	18.3\\
			& 	356		& 	62.0\\
			& 	384		& 	9.8\\
\hline
$^{60}Co$ 	&	1173 	&	99.8 \\
			& 	1332		&	100 \\
\hline
$^{137}Cs$ 	& 	32		& 	5.6\\
		 	& 	662		&	85.1 \\
\hline
\end{tabular}
\end{table}

Plateau measurement have been used for this since in order to measure a pulse height spectrum, either the threshold needs to be set extremely low, or a high bias voltage needs to be used. Neither option is practical with the amplifiers that were available, which are, of course, optimized for the much larger neutron signals. At a typical gas amplification used in neutron detection, the threshold cannot be set low enough to study $\gamma$-ray signals due to electronic noise. Increasing the bias voltage results in a high gas gain but in turn leads to poor energy resolution. 

\subsubsection{Low- versus high-energy $\gamma$-rays}

Both detectors become sensitive to low-energy photons at lower gain, but potentially are more sensitive to the high-energy photons when the $\gamma$ plateau is reached. This is visible in the comparison between the measurements of two sources in figure~\ref{fig:sources}. The point-like sources were placed directly on he detector vessels. Counts are shown for the tube ($B_4C$-lined or $^3He$-filled) closest to the source. The $^{133}Ba$ source with x-rays at 33~keV and $\gamma$-rays of 80~keV and several close to 300~keV can be seen as a low energy source, whereas $^{60}Co$ emits only $\gamma$-rays above 1~MeV. We find that as the bias voltage is increased the Barium source generates more counts at first. This corresponds to interactions in the gas as well as the walls where photo- or Compton electron energies are low and specific energy deposit relatively high. As the voltage is increased further, the count rate due to the Cobalt source exceeds that of Barium since now a larger number of interactions, primarily in the solid elements, contribute with higher energy electrons which have lower specific energy loss. 

In a realistic configuration where a gas detector is set up to detect neutrons, the energy threshold should be set so that the signals from photons are rejected while those from neutrons are not. No sharp cut-off for the highest pulse hight resulting from a specific primary particle energy exists due to the statistical nature of gas amplification and charge transport. The final contamination due to $\gamma$-ray signals will be due to those events where a $\gamma$ signal including statistical fluctuation is over the threshold. This is most likely to be due to a low-energy $\gamma$-ray (since the specific energy loss is then high), such as 60~keV. At the first glance this fact is encouraging since it is much easier to shield low-energy photons. Note however, that interactions of high-energy photons as well as nuclear reactions, such as neutron capture or decay of activated materials, often result in emission of x-rays and internal conversion electrons which have just the energy that is most likely to contribute to background. It is therefore important to carefully consider the external radiation environment as well as the secondary sources that may exist in the immediate vicinity of the detectors. The details of the optimization for shielding for this purpose may differ from that for traditional radioprotection purposes.

\subsubsection{High statistics measurement}

\begin{figure}[tbp] 
\centering
\includegraphics[width=1.05\textwidth]{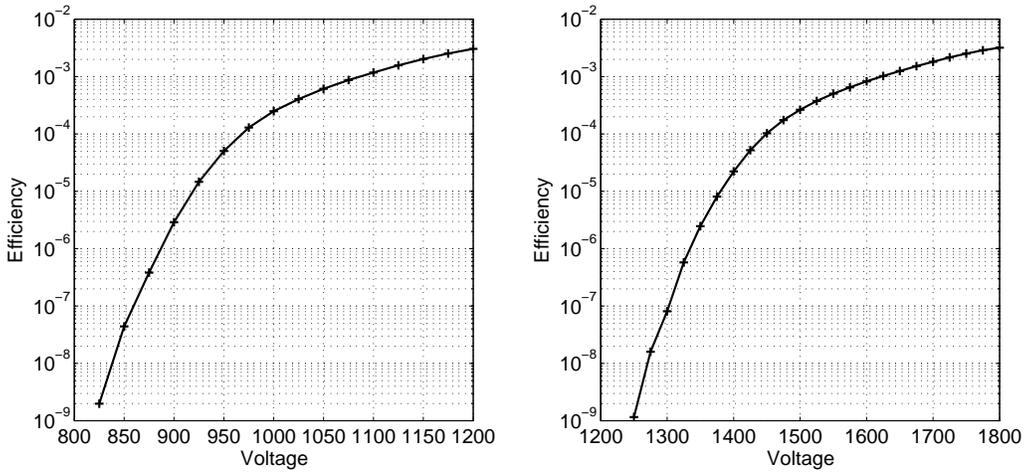}
\caption{Plateau measurement with the Multi-Grid $^{10}B$ detector (left) and a Multi-Tube $^3He$ detector (right) with a 164~MBq $^{137}Cs$ source.}
\label{fig:sources2}
\end{figure}

Figure~\ref{fig:sources2} shows a measurement with a much stronger $^{137}Cs$ source. The setup was shielded with polyethylene, $B_4C$ and lead. The source was in this case not in direct contact with the detector and the counts are measured for 20 tubes in the front of the B-10 detector and in all 37 tubes of the He-3 detector. This represents an approximately equal solid angle with respect to the source. Efficiency values have been normalized and are shown per tube. In this case, counts due to the source could be clearly distinguished over the background for all operating voltages. The statistical error bars are smaller than the resolution of the plot and are not shown. The uncertainty does, however, originate from the true rate of $\gamma$-rays incident on the detector. The main factors are the solid angle of the detector in the field of view of the source and distribution and self-absorption of the source (which is roughly cylindrical). These parameters are constant for all points in each plot and represent an uncertainty in the overall scale of the curves that is not higher than factor 2. 

Results in figures~\ref{fig:sources} and \ref{fig:sources2} show that, as expected, the nature of the neutron converter does not influence the response to $\gamma$-rays. For both detectors, a sensitivity on the order of $10^{-9}$ per tube is found for the $^{137}Cs$ source for reasonable operation settings. Both detectors would have an efficiency on the order of 1\% (over the whole detector) if they were to be operated as $\gamma$ detectors. 



%
%

\subsection{In-beam Measurement}

A new version of the Multi-Grid detector has been built in a configuration that can be mounted in the IN6 time-of-flight chopper spectrometer at the ILL. This detector contains 96 frames in 6 assemblies of 16 frames, resulting in an active area of $0.15 m^2$. The detector replaces 25 $^3He$ tubes of the standard compliment of IN6. 

The main goal of the tests on IN6 has been the characterization of neutron detection side-by-side with conventional $^3He$ detectors, and these results will be presented elsewhere. For our purpose here, it is interesting to note that in a chopper spectrometer, neutrons arriving at the detector show a time structure. Those scattered elastically in the sample form a distinct peak in the time spectrum, while those scattered inelastically arrive earlier or later than the elastic peak (depending on the energy transfer to or from the neutron). In particular, it is possible to chose a sample which will only create an elastic signal with essentially no neutrons arriving at other times.

As already alluded to in the introduction, a large $\gamma$-ray background is generated by the instrument itself as well as the surrounding equipment. The background originating in the instrument also shows a time structure. When the chopper is open, a neutron pulse starts traversing the following beam line elements. There are collimators, sample environment and sample and finally a beam stop. Each component generates $\gamma$-rays due to the neutrons absorbed, which in turn travel with the speed of light in a random direction. This flux was measured using a $NaI$ scintillator detector placed next to the $^{10}B$ prototype, showing that on the order of 1 million $\gamma$-rays per second were passing through the area of the prototype. The total rate measured with the prototype can be as low as a few Hz, corresponding to the neutron flux.

\begin{figure}[tbp] 
\centering
\includegraphics[width=1\textwidth]{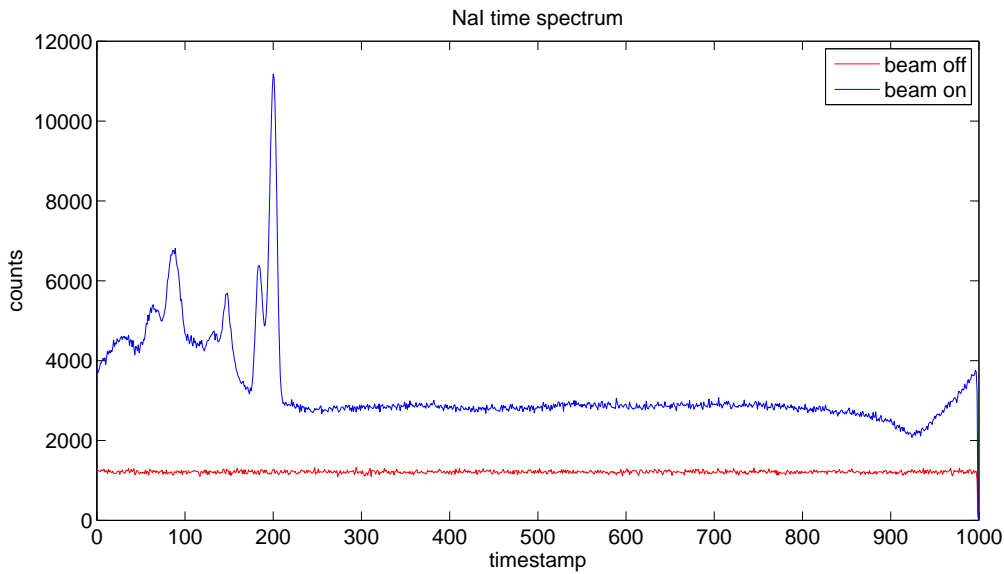}
\caption{Time spectrum of the $NaI$ $\gamma$-ray detector. The peaks correspond to photons generated at a fixed time delay after the chopper opening. The flat level is the constant background generated during the chopper closed state as well as all other sources in the guide hall.}
\label{fig:nai}
\end{figure}

\begin{figure}[tbp] 
\centering
\includegraphics[width=1\textwidth]{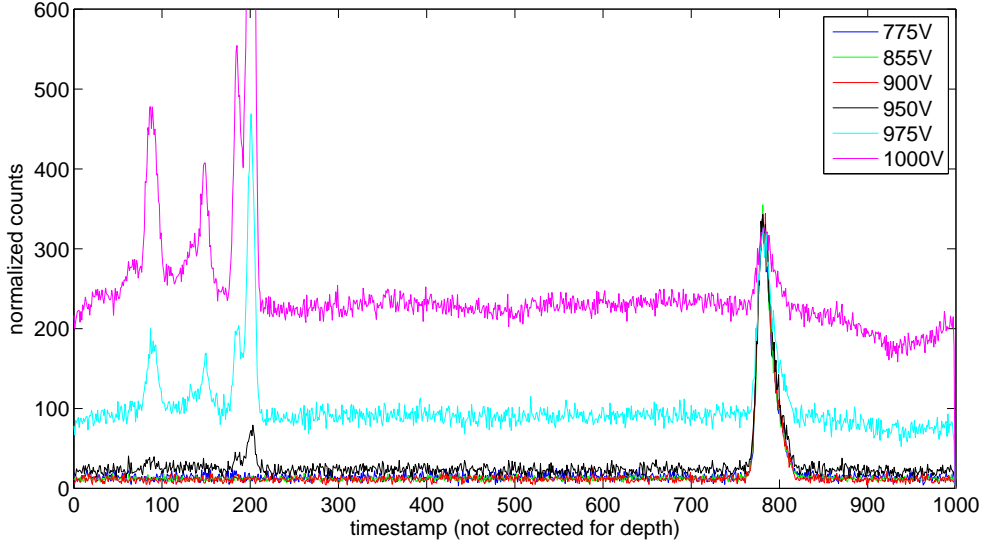}
\caption{Time spectrum of the $^{10}B$ prototype for a range of bias voltages. No evidence of the $\gamma$-peaks is visible until the voltage reaches 950V. The peak at the channel numbers 770-810 is the elastic neutron peak. Note that no timing correction for the depth of the detector was performed here, since it cannot be done in a consistent way for both $\gamma$ and $n$ at the same time -- and here we are interested in $\gamma$ -- therefore the neutron peak appears wider than it normally would.}
\label{fig:b10time}
\end{figure}

Figure~\ref{fig:b10time} shows the corresponding time spectrum of the $^{10}B$ detector prototype for several bias voltages. We now find the peak corresponding to the elastic scattering of neutrons in the sample. Peaks corresponding to $\gamma$ are only present in the measurements at far higher bias voltage than that chosen for normal operation (850V). The rate of neutron detection was $4~Hz$ in this measurement, with a neutron efficiency determined to be 55\% at 2.5{\AA}~\cite{cite:correa}.

\begin{figure}[tbp] 
\centering
\includegraphics[width=0.4\textwidth]{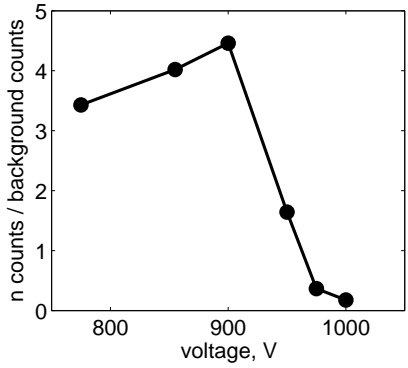}
\caption{Evolution of the ratio of the neutron signal in the elastic peak to background corresponding to the largest $\gamma$ peak.}
\label{fig:ptb}
\end{figure}

Let us define a peak-to-background ratio by selecting the neutron peak, channels 770-810 and the $\gamma$ background in channels 185-210, i.e. when there is a known $\gamma$ peak. Figure~\ref{fig:ptb} shows the evolution of this ratio as the bias voltage is increased. We see that for the first 3 voltages, the ratio is increasing, indicating that the detector gains neutron efficiency, while not becoming sensitive to the gamma flux (which is at least 5 orders of magnitude higher). Raising the voltage further, drastically affects the ratio since the $\gamma$ signals can now be detected over the threshold. 

\subsection{Trade-off between neutrons and gamma}

\begin{figure}[tbp] 
\centering
\includegraphics[width=.7\textwidth]{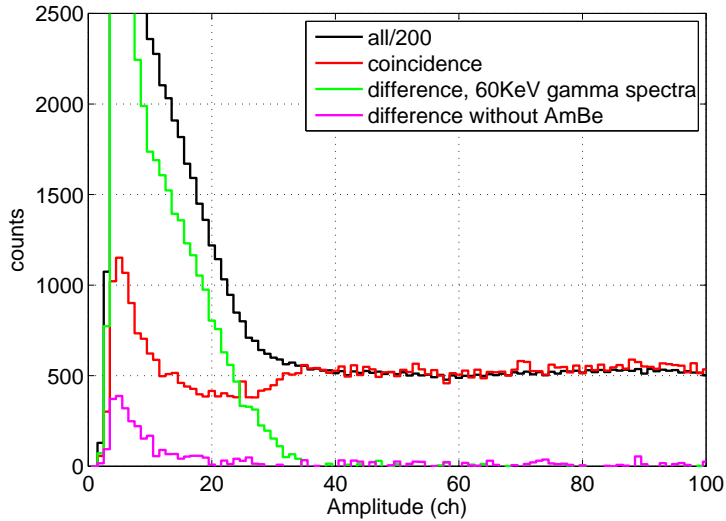}
\caption{Energy spectrum of neutron events compared to $\gamma$-ray spectra for the $^{10}B$ detector.}
\label{fig:spec}
\end{figure}

The energy spectrum of neutron conversion fragments that reach the gas has no lower limit for a solid film detector. Therefore any lower level threshold will reject some neutron events. A minimum required threshold is determined by the end-point of the $\gamma$ spectrum. A comparison of neutron and gamma spectra are shown in figure~\ref{fig:spec}. 

In order to perform this measurement, a thin detector with only one boron layer was illuminated with both a neutron beam and a $^{241}Am$ source simultaneously. In this way comparable neutron and $\gamma$ counts can be obtained (whereas in case of an $AmBe$ source only, the gamma counts greatly exceed neutron counts for a high enough gas gain). A $NaI$ detector shielded with lead was placed close to the boron layer in order to measure the $478 keV$ $\gamma$-rays emitted in 94\% of the neutron conversions. The B-10 and the $NaI$ detectors were readout by the same data acquisition system and events were assigned a timestamp. This makes it possible to examine both singles and coincident spectra in the same measurement. The total rate in the $^{10}B$ detector was 3.3~kHz and the rate of the $NaI$ with the gate around the 478~keV peak was 190~Hz. The rate of random coincidences is therefore expected to be negligible. The coincidence rate was 11~Hz.

The B-10 detector was operated at a very high gain in order to make it sensitive to photon events. The neutron spectrum extends to much larger energies and only its lower part can be measured without saturation in these conditions. Selecting the $478 keV$ peak in the $NaI$ in coincidence with a signal in the B-10 detector allows to identify true neutron conversion events, even those whose energies normally make them indistinguishable from $\gamma$. These are shown in red in the figure. The probability to detect a $478 keV$ photon emitted from the boron layer is of course much smaller than one. However, since we know that for large energies essentially only neutrons contribute to the spectrum, scaling allows to match the coincident spectrum with the total spectrum. The difference then corresponds to the spectrum due to all non-neutron events as shown in green. Finally, verifying that most of this difference spectrum vanishes when the $\gamma$ source is removed, confirms that it corresponds to the $\gamma$ spectrum of $^{241}Am$ -- predominantly $60 keV$.

\begin{figure}[tbp] 
\centering
\includegraphics[width=.7\textwidth]{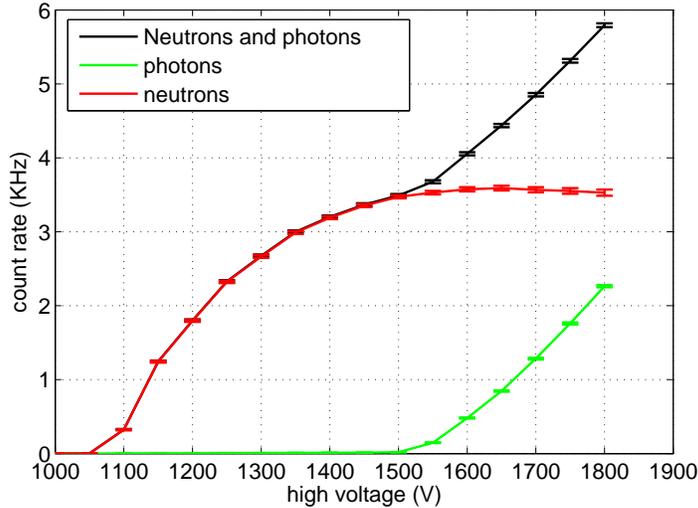}
\caption{Plateau measurement for neutrons and $\gamma$-rays spectra for the $^{10}B$ detector.}
\label{fig:plateau}
\end{figure}

The fraction of neutrons rejected due to the overlap of the $\gamma$ and neutron spectra can be estimated in a plateau measurement, see figure~\ref{fig:plateau}. Setting the high voltage at the point just before the rise of the $\gamma$ detection results in approximately 7\% fewer neutron counts compared to the maximum of the plateau. Note however, that a certain threshold level is required also to reject electronic noise, and so the \emph{loss} compares to all neutron events where some part of the energy is deposited in the gas, not to those event that can be detected over the electronic noise level -- which will be a smaller fraction that becomes zero for the same amplitude of $\gamma$ signals and electronic noise.

\section{Conclusions}

The discrimination between neutron and photon signals presents a challenge in many types of neutron detectors. In the field of neutron scattering science this is particularly true due to relatively low scattering rates and potentially very high $\gamma$ backgrounds. A high level of discrimination can be reached with the conventional $^3He$ tube. As the search for alternatives for the the waning supply of $^3He$ began, concerns have been high for the $\gamma$ rejection achievable with the new technologies. Over the course of tests of the $^{10}B$ thin film detectors, we have found that the $\gamma$ rejection need not be lower in these detectors than in $^3He$ tubes.

With this paper, we have investigated in depth the physical effects and geometric considerations that affect the sensitivity to $\gamma$-rays in gas-based detectors for thermal neutrons. GEANT4 simulations have been used to investigate the effects of photon energies, gas pressure and the contribution of solid and gaseous parts of the detector to the $\gamma$ detection probability. We find that an energy threshold of at most $100 keV$ is sufficient to reject all photons. Those photons that deposit larger amount of energy are only those where energy is spread over multiple detector cells or tubes and can be rejected by data acquisition. Our experience with $^{10}B$ prototypes tested in beam and with sources agree with these findings.

While measurements and simulations presented here were done for the $^{10}B$ Multi-Grid detector, the underlying physical processes responsible for $\gamma$ counts in a neutron detector will be very similar for other neutron converters, such as solid layers of $^6Li$ or $^3He$ and $^{10}BF_3$ gases, provided that gas-based readout is used. We therefore hope that this work will aid the efforts to develop such technologies, as well as to extend the investigation of $\gamma$-ray sensitivity to other classes of detectors for thermal neutrons.

%
%
%
%
%

\acknowledgments

We would like to thank M.~M.~Koza, M.~Zbiri and J.~Halbwachs for the help with measurements at IN6; C.~H\"oglund, J.~Birch and L.~Hultman for the contribution of the boron layers for the prototype. The work has been supported by the CRISP project (European Commission 7th 
Framework Programme Grant Agreement 283745).

\end{document}